\documentclass[conference, twocolumn, 9pt]{IEEEtran}	

\PassOptionsToPackage{utf8}{inputenc}
\usepackage{inputenc}

\PassOptionsToPackage{T1}{fontenc} %
\usepackage{fontenc}
\PassOptionsToPackage{dvipsnames}{xcolor}
\usepackage{soul} %

\PassOptionsToPackage{fleqn}{amsmath}       %
\usepackage{amsmath}
\usepackage{mathtools}
\usepackage{xfrac}
\usepackage[]{siunitx}
\DeclareSIUnit\torr{Torr}
\DeclareSIUnit\atoms{atoms}

\usepackage[inline]{enumitem}

\usepackage[vlined,ruled,linesnumbered,longend]{algorithm2e}
\usepackage{graphicx}
\usepackage{textcomp}
\usepackage{xspace}
\usepackage[size=tiny]{todonotes}
\PassOptionsToPackage{hyphens}{url}
\usepackage[hidelinks]{hyperref}

\usepackage{subfig}
\usepackage{multirow}
\usepackage{multicol}
\usepackage[export]{adjustbox}
\usepackage{float}
\usepackage{booktabs}
\usepackage{cite}
\usepackage{balance}

\usepackage{bm} %

\newcommand{\ie}{i.\,e.\xspace}

\newcommand{\eg}{e.\,g.\xspace}

\newcommand{\etal}{\textit{et~al}.\xspace}

\newcommand{\fiction}{\emph{fiction}\xspace}
\newcommand{\siqad}{\emph{SiQAD}\xspace}

\newcommand{\hsi}{H-Si(100)-2\texttimes1\xspace}

\newcommand{\refsec}[1]{Section~\ref{sec:#1}}
\newcommand{\reffig}[1]{Fig.~\ref{fig:#1}}

\newcommand{\reftab}[1]{Table~\ref{tab:#1}}

\begin{document}

\title{Design Automation for Silicon Dangling Bond Logic under Atomic Defects on the \hsi Surface}
\title{A Physical Design Flow for Silicon Dangling Bond Logic under Atomic Defects on the \hsi Surface}
\title{Defect-Aware Physical Design of Silicon Dangling Bond Logic}
\title{Atomic Defect-Aware Physical Design of \\ Silicon Dangling Bond Logic on the \hsi Surface}

\author{\IEEEauthorblockN{Marcel Walter\IEEEauthorrefmark{1}, Jeremiah Croshaw\IEEEauthorrefmark{2}\IEEEauthorrefmark{3}, Samuel Sze Hang Ng\IEEEauthorrefmark{4}, Konrad Walus\IEEEauthorrefmark{4}, Robert Wolkow\IEEEauthorrefmark{2}\IEEEauthorrefmark{3}, and Robert Wille\IEEEauthorrefmark{1}\IEEEauthorrefmark{5}} \\[-1ex]
	\IEEEauthorblockA{\IEEEauthorrefmark{1}Chair for Design Automation, Technical University of Munich, Germany \\
		Email: \{\href{mailto:marcel.walter@tum.de}{marcel.walter}, \href{mailto:robert.wille@tum.de}{robert.wille}\}@tum.de} \\[-2ex]
	\IEEEauthorblockA{\IEEEauthorrefmark{2}Department of Physics, University of Alberta, Edmonton, Canada \\ %
		Email: \{\href{mailto:croshaw@ualberta.ca}{croshaw}, \href{mailto:rwolkow@ualberta.ca}{rwolkow}\}@ualberta.ca} \\[-2ex]
	\IEEEauthorblockA{\IEEEauthorrefmark{3}Quantum Silicon Inc., Edmonton, Canada} \\[-2ex]
	\IEEEauthorblockA{\IEEEauthorrefmark{4}Department of Electrical and Computer Engineering, University of British Columbia, Vancouver, Canada \\ %
		Email: \{\href{mailto:samueln@ece.ubc.ca}{samueln}, \href{mailto:konradw@ece.ubc.ca}{konradw}\}@ece.ubc.ca} \\[-2ex]
	\IEEEauthorblockA{\IEEEauthorrefmark{5}Software Competence Center Hagenberg GmbH, Austria}}

\maketitle

\begin{abstract}
	Although fabrication capabilities of \emph{Silicon Dangling Bonds} have rapidly advanced from manual labor-driven laboratory work to automated manufacturing in just recent years, sub-nanometer substrate defects still pose a hindrance to production due to the need for atomic precision. In essence, unpassivated or missing surface atoms, contaminants, and structural deformations disturb the fabricated logic or prevent its realization altogether. Moreover, design automation techniques in this domain have not yet adopted any defect-aware behavior to circumvent the present obstacles. In this paper, we derive a surface defect model for design automation from experimentally verified defect types that we apply to identify sensitivities in an established gate library in an effort to generate more robust designs. Furthermore, we present an automatic placement and routing algorithm that considers scanning tunneling microscope data obtained from physical experiments to lay out dot-accurate circuitry that is resilient against the presence of atomic surface defects. %
	This culminates in a holistic evaluation on surface data of varying defect rates that enables us to quantify the severity of such defects. We project that fabrication capabilities must achieve defect rates of around $\SI{0.1}{\percent}$, if charged defects can be completely eliminated, or $< \SI{0.1}{\percent}$, otherwise. This realization sets the pace for future efforts to scale up this promising circuit technology.
\end{abstract}

\section{Introduction \& Motivation}

With the decline of Moore's Law, research has turned to alternative circuit technologies in the search for promising post-CMOS candidates. A relatively new contestant in this domain are \emph{Silicon Dangling Bonds}~(SiDBs) that act as atomically-sized quantum dots and that have seen tremendous fabrication advancements in the recent years~\cite{haider2009controlled, huff2017atomic, pavlicek2017tip, achal2018lithography, Wolkow14, D0NR08295C, wyrick2019atom}. Under the term \emph{atomic silicon quantum dots}, their application for the creation of nanometer-sized logic cells has been investigated which led to the successful demonstration of a \mbox{sub-\SI{30}{\square\nm}} SiDB OR gate and wire segments on hydrogen-passivated silicon surfaces~\cite{huff2018binary}. Relying on Coulomb interaction instead of the transmission of electric current, SiDB logic implements the \emph{Field-coupled Nanocomputing}~(FCN) paradigm~\cite{Anderson14} that offers logic-in-memory devices~\cite{jiao2003building, qi2003molecular} and promises energy dissipation capabilities below the \emph{Landauer limit}~\cite{landauer1961irreversibility, keyes1970minimal, lent2006bennett, jiao2003building, toth1999quasiadiabatic}, or clock frequencies in the terahertz regime~\cite{Timler02, livadaru2010danglingbond, ng2020thes, chiu2020poissolver, chiu2020thes}.

Motivated by this, the research community has already taken interest in the SiDB platform for logic design; an effort that resulted in the creation of various physical simulators~\cite{ng2020siqad, chiu2020poissolver, ng2020thes, drewniok2023quicksimIEEE, drewniok2023temperatureIEEE, drewniok2023need}, various manually designed circuits and some gate libraries~\cite{ng2020siqad, ng2020thes, chiu2020thes, bahar2020atomic, vieira2021novel, vieira2022three, walter2022hexagons, ng2023ablueprint}, as well as algorithms for placement and routing of SiDB gates~\cite{walter2022hexagons, hofmann2023hexagonalization}.

However, SiDB fabrication requires atomic precision and, thus, is prone to substrate defects at the sub-nanometer level. These defects naturally occur during the substrate preparation, \ie, in the process of preparing a pure hydrogen-passivated silicon \mbox{(\hsi)} surface for SiDB fabrication.  They are generally classified as any atomic structure that does not follow the \hsi surface reconstruction, in which each surface silicon atom is bonded to a neighboring silicon atom creating a dimer pair, a hydrogen atom from the passivation, and two silicon atoms in the bulk of the crystal~\cite{boland1993}.  Such defects could include unpassivated surface silicon atoms, missing silicon atoms, contaminant atoms, or structural deformations (which is covered in more detail in \refsec{defects}). 

Current fabrication of SiDB logic necessitates the scanning for defect-free regions that are large enough to host the intended layout on the substrate at hand~\cite{rashidi2020deep}. While being a useful proof-of-concept demonstration, this fabrication approach is not only wasteful, but also increasingly unrealistic with growing layout size as it can accommodate for a handful of gates at the utmost. Hence, the presence of surface defects imposes a hindrance on scaling SiDB design size. At the same time, operating at the atomic-scale, material defects are largely common at the current fabrication capabilities.

Accordingly, the electrostatic effects of such surface defects have been closely examined---providing insights into how they might limit SiDB fabrication and the operation of fabricated SiDB devices~\cite{croshaw2020atomic, huff2019landscape}. This work leverages these defect analysis findings to enable automatic layout design under the presence of atomic surface defects.

In this regard, the paper at hand proposes the following contributions:
\begin{enumerate}
	\item the derivation of an atomic defect model by the establishment of equivalence classes among 13 experimentally verified \mbox{\hsi} defect types to guide design automation methodologies,
	\item a case study of applying said defect model to the established \emph{Bestagon} gate library~\cite{walter2022hexagons}, which led to the identification of sensitivities of some gates to certain defects for which we propose more robust redesigns,
	\item an automatic placement and routing algorithm that considers real \emph{Scanning Tunneling Microscope}~(STM) surface scans obtained from physical experiments as well as simulated surface data to design functioning dot-accurate SiDB circuit layouts in the presence of atomic surface defects by avoiding disturbed regions, and
	\item a culmination of the previously mentioned contributions into a holistic experimental evaluation on \hsi surfaces of variable defect rates that quantifies the severeness of---particularly---charged atomic defects.
\end{enumerate}

\urlstyle{tt}

Thereby, we propose the first defect-aware framework for SiDB logic that amalgamates fabrication and design automation. Experimental evaluations on both real and simulated \hsi surfaces allow us to estimate the required surface manufacturing quality in terms of the defect rate for large-scale SiDB device fabrication in future efforts to be around \SI{0.1}{\percent}, if charged defects can be completely eliminated or $< \SI{0.1}{\percent}$, otherwise.%

The remainder of this manuscript is structured as follows: in an effort to establish this paper as a stand-alone work, \refsec{prelims} reviews related material on SiDB fabrication and their logic platform to constitute the foundation upon which this paper is built. Afterward, \refsec{defects} introduces atomic defects on the \mbox{\hsi} surface and discusses their effects on SiDB systems. Based on that, \refsec{proposed} presents the proposed defect-aware physical design methodology by first establishing a surface defect model and, then, discussing algorithmic details. An experimental evaluation of the approach is conducted in \refsec{eval}. Finally, \refsec{concl} concludes the paper and gives an outlook on future work in the domain.

\section{Silicon Dangling Bond Logic} \label{sec:prelims}

As an implementation of the FCN paradigm~\cite{Anderson14}, the utilization of SiDBs has recently gained momentum. Using an \mbox{atomically-sharp} tip of a \emph{Scanning Tunneling Microscope}~(STM), individual dangling bonds can be created on a hydrogen-passivated silicon surface at the single-atom scale~\cite{pavlicek2017tip, huff2017atomic, achal2018lithography, rashidi2022automated, huff2019electrostatic, achal2019detecting, onoda2021ohmic, altincicek2022atomically}. These dangling bonds act as quantum dots and are used to represent logic states and to realize Boolean operations at the limit of physical scaling~\cite{haider2009controlled, huff2018binary, Wolkow14, pitters2011charge, rashidi2018initiating}.

\begin{figure}[t!]
	\centering
	\subfloat[Electron injection above the hydrogen atom.]{
		\includegraphics[width=.4\linewidth]{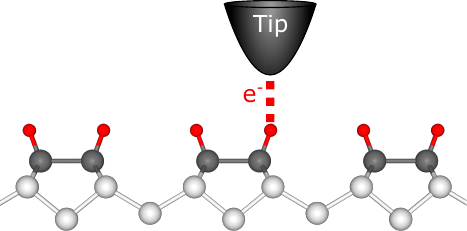}
		\label{fig:sidb-creation_a}
	} \hfil
	\subfloat[Hydrogen desorption from the surface.]{
		\includegraphics[width=.4\linewidth]{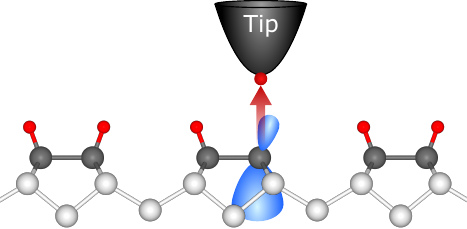}
		\label{fig:sidb-creation_b}
	} \\
	\subfloat[Newly created SiDB.]{
		\includegraphics[width=.4\linewidth]{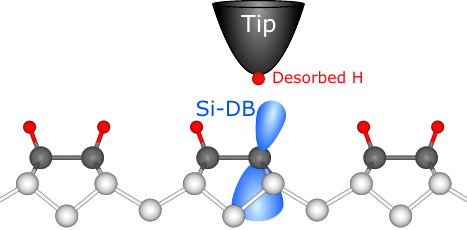}
		\label{fig:sidb-creation_c}
	} \hfil
	\subfloat[\hsi surface lattice aligned with an STM image.]{
		\includegraphics[width=.4\linewidth]{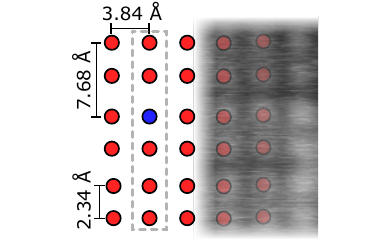}
		\label{fig:sidb-creation_d}
	} 
	\caption{Illustration of the SiDB fabrication process on the \mbox{\hsi} surface using an STM tip. The atomic surface structure is depicted as a side-view ball-and-stick model.}
	\label{fig:sidb-creation}
\end{figure}

The most commonly used surface phase for SiDB creation is \mbox{\hsi} whose atomic structure is illustrated in \reffig{sidb-creation}. The surface consists of discretely defined sites (shown as red atoms in \reffig{sidb-creation_d}), where SiDBs can be fabricated with atomic precision.  By using the scanning probe tip to inject current into the H-Si bond, it is possible to selectively remove single hydrogen atoms from the surface, leaving behind an SiDB (shown in blue in \reffig{sidb-creation_b}--\ref{fig:sidb-creation_d}).

The resulting SiDBs may possess 0, 1, or 2 electrons, corresponding to positive, neutral, and negative charge states, respectively.\footnote{Note that, in the following only negative or neutral dangling bonds are of interest. Positively charged ones are not relevant for gate configurations~\mbox{\cite{ng2020siqad, huff2018binary}}} The corresponding charge states can be controlled by environmental factors such as the bulk dopant concentration~\cite{rashidi2016timeresolved} and the presence of electric fields~\cite{pitters2011charge,huff2019electrostatic}. A groundbreaking demonstration by Huff \etal has experimentally demonstrated that careful configurations of pairs of SiDBs can be used to realize logic components~\cite{huff2018binary}.  These SiDB pairs are observed to share a single additional electron between them which can be manipulated to occupy the left or right SiDB of the pair indicating a binary 0 or 1 state; a behavior that was coined \emph{Binary-Dot Logic}~(BDL)~\cite{huff2018binary}.

\begin{figure}[t!]
	\centering
	\includegraphics[width=.8\linewidth]{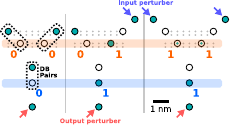}
	\caption{Conceptual recreation of the SiDB OR gate proposed and fabricated on the \hsi surface by Huff \etal~\cite{huff2018binary}.}
	\label{fig:lattice-and-bdl}
\end{figure}

These electronic properties enable the realization of BDL wire structures as well as a sub \SI{30}{\square\nm} logic OR gate%
~\cite{huff2018binary}. A simulated reproduction of the OR gate using the SiDB simulator \siqad~\cite{ng2020siqad} is shown in \reffig{lattice-and-bdl}. Here, %
the input bit states are set by the addition of a peripheral SiDB, dubbed a \textit{perturber}~\cite{huff2018binary, ng2020siqad}, which exerts an external field %
on the input SiDB pair to emulate the presence of an input BDL wire at the logic $1$ state. %
When one or both of the input SiDB pairs are set to logic $1$ by input perturbers, the output also toggles to logic $1$ as expected of an OR gate. It is to be noted that the need for these perturbers will be alleviated upon the future development of I/O devices.

The recent physical accomplishments and the upcoming commercialization~\cite{Wolkow2021, Wolkow2021a} in the domain of SiDB logic have triggered an increasing interests in design automation methods for this technology---yielding first CAD tools, circuit layouts, and physical design algorithms~\cite{ng2020siqad, chiu2020poissolver, ng2020thes, drewniok2023quicksimIEEE, drewniok2023temperatureIEEE, drewniok2023need, chiu2020thes, bahar2020atomic, vieira2021novel, vieira2022three, walter2022hexagons, ng2023ablueprint, hofmann2023hexagonalization}. However, none of these methodologies and proposals take into account that the fabrication of sub-nanometer structures is naturally error-prone and that atomic defects of the substrate are inherent to this endeavor and will continue to be for the foreseeable future~\cite{croshaw2020atomic, huff2019landscape}. The following section covers common atomic defects on the \hsi surface and their influence on SiDB logic.

\section{Atomic Surface Defects} \label{sec:defects}

Despite the relative cleanliness of \hsi compared to other crystal faces of silicon~\cite{boland1993}, there still exists a natural concentration of defects that cannot be completely avoided with current \emph{in-situ} preparation methods. These defects can be broadly described as any collection of atoms in the crystal that do not form the 2$\times$1 surface phase, where each surface silicon atom is host to only one hydrogen. These could include unpassivated surface silicon atoms, missing silicon atoms, contaminant atoms, or structural deformations. To this end, defects prevent the creation of atomically identical SiDBs due to their varying structures. Additionally, the proximity of defects alters SiDB behavior and, consequently, corrupts implemented logic gates.

\begin{figure}[!t]
	\centering
	\subfloat[STM surface scan of \SI{19}{\nano\meter} $\times$ \SI{18}{\nano\meter} with visible defects.]{
		\includegraphics[width=.8\linewidth]{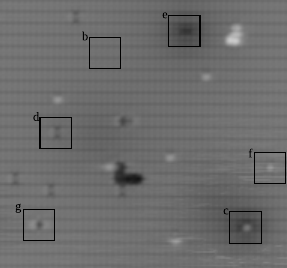}
		\label{fig:defects:surface}
	} \\
	\subfloat[Defect-free \hsi.]{
		\includegraphics[width=.45\linewidth]{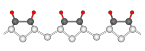}
		\label{fig:defects:hsi}
	} 
	\subfloat[Dangling bond.]{
		\includegraphics[width=.45\linewidth]{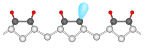}
		\label{fig:defects:db}
	} \\
	\subfloat[Dihydride pair.]{
		\includegraphics[width=.45\linewidth]{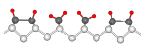}
		\label{fig:defects:dihydride}
	} 
	\subfloat[Silicon vacancy.]{
		\includegraphics[width=.45\linewidth]{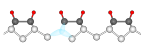}
		\label{fig:defects:vacancy}
	} \\
	\subfloat[Siloxane dimer.]{
		\includegraphics[width=.45\linewidth]{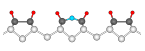}
		\label{fig:defects:siloxane}
	} 
	\subfloat[Missing dimer.]{
		\includegraphics[width=.45\linewidth]{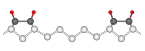}
		\label{fig:defects:missing_dimer}
	} \\
	\caption{A \hsi surface and common atomic defects found thereon depicted as side-view ball-and-stick models.}
	\label{fig:defects}
\end{figure}

As an example, \reffig{defects:surface} depicts an empty states STM scan acquired at \SI{1.3}{\volt} and \SI{50}{\pico\ampere} of a physically fabricated \hsi surface. The black frames with alphabetical labels indicate different atomic defects. \reffig{defects:hsi} to \reffig{defects:missing_dimer} illustrate these defects as a side-view ball-and-stick model. The following list gives a brief explanation of their nature:\footnote{A more detailed discussion of these and other commonly occurring defects can be found in~\cite{croshaw2020atomic}.}

\begin{figure}[!t]
	\centering
	\includegraphics[width=.8\linewidth]{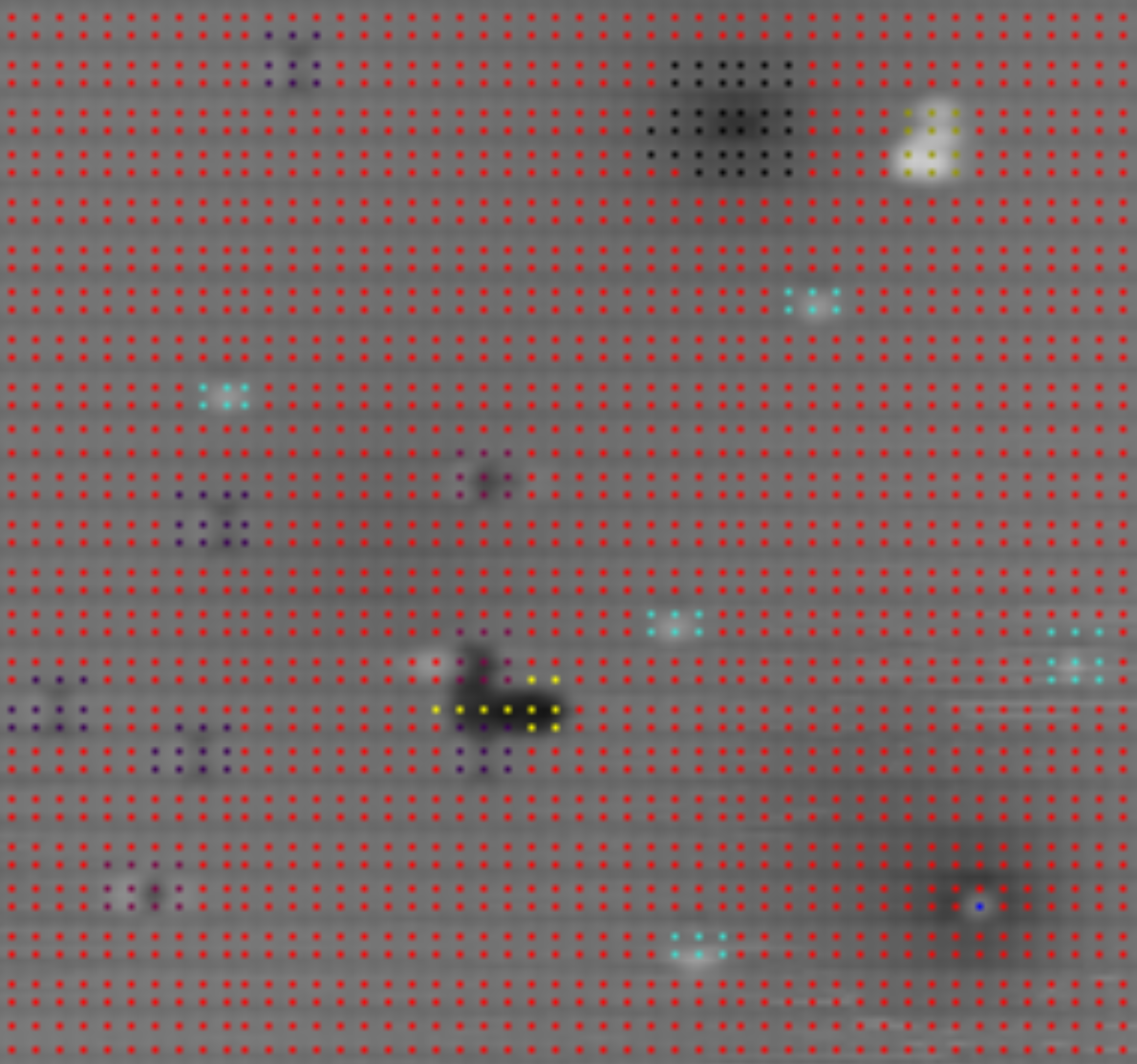}
	\caption{The same STM image as shown in \reffig{defects:surface} with a lattice overlay. Color key: \emph{red}: defect-free H-Si, \emph{blue}: dangling bond, \emph{black}: silicon vacancy, \emph{purple}: dihydride pair, \emph{magenta}: missing dimer, \emph{light blue}: siloxane and similar, \emph{yellow}: etch pit (multiple missing dimers), \emph{gold}: unknown.}
	\label{fig:defect_lattice_figure}
\end{figure}

\begin{itemize}
	\item[\ref{fig:defects:hsi}] The defect-free \hsi surface phase.  Each surface silicon atom in dark gray is paired with another surface silicon atom creating a dimer pair.  Each silicon atom of a dimer is then passivated with a single hydrogen atom.  Each silicon atom in this configuration is capable of hosting a single, chemically identical dangling bond.
	
	\item[\ref{fig:defects:db}] A silicon atom that is not terminated with hydrogen leaving a dangling bond. These can be intentionally created, or found from incomplete hydrogen passivation.
	
	\item[\ref{fig:defects:dihydride}] A dihydride pair, where no dimer bond forms, leaving each silicon atom to bond with two hydrogen atoms. Dihydrides are found more often when the crystal annealing temperature is too low~\cite{Boland1992}. 
	
	\item[\ref{fig:defects:vacancy}] A silicon vacancy, where a single silicon atom is missing from the lattice leaving unsatisfied subsurface dangling bonds.  
	
	\item[\ref{fig:defects:siloxane}] A siloxane dimer, which features a single oxygen atom between the two silicon atoms of the dimer. This defect occurs in high concentration when the preparation chamber is contaminated with water molecules. 
	
	\item[\ref{fig:defects:missing_dimer}] A missing dimer where both silicon atoms are absent.
\end{itemize}

Atomic defects can furthermore also appear in various combinations on the same or adjacent dimers which increases their affects on logic placed in their vicinity~\cite{croshaw2020atomic}. %

To the best of the authors' knowledge, no related work on SiDB logic and design automation has considered the impact atomic defects have on their proposed layouts. This disregard of physical effects leaves most approaches conflicting with existing fabrication capabilities.

In the following section, we are addressing this shortcoming by proposing a physical design methodology that is aware of atomic defects. Consequently, the layouts automatically generated by the proposed approach avoid defective surface positions. %

\section{Defect-Aware Physical Design} \label{sec:proposed}

In this section, we introduce a solution that addresses the shortcomings of existing physical design methods for SiDB logic discussed above.
To this end, we first propose an abstract surface defect model for automatic design that is based on the physical properties of the identified atomic surface defects. Afterward, we propose a \mbox{defect-aware} physical design method that, utilizing the proposed model, is able to realize SiDB logic that behaves as intended on an otherwise defective surface.

\subsection{Surface Defect Model} \label{sec:defect_model}

Before being able to apply any defect-aware design methodology, the atomic structure of the specific surface at hand must be analyzed and defects identified. In this instance, defects are autonomously classified using a \emph{Convolutional Neural Network}~(CNN) similar to that developed in~\cite{rashidi2020deep}. Its input is an STM image like that in \reffig{defects:surface}. The CNN provides a pixel-based classification of the surface corresponding to the defect types as its output. The classification is then correlated to the lattice positions of the \hsi surface yielding a coordinate-based defect assignment as shown in \reffig{defect_lattice_figure}. Hydrogen-terminated silicon (defect-free dimers) is labeled in red, while various defects are labeled as described in the caption.\footnote{Looking at the defect lattice positions, one can see that some assigned labels extend farther than the underlying physical defect.  This is due to the output of the CNN.  When labeling the training data, defects were marked based on their contrast in the acquired images and not the exact lattice coordinates.  Since some defects had a strong influence on the contrast of adjacent, defect-free atoms, they were labeled as the same defect to avoid any aliasing in defect classification.  It is possible to apply a filter as a post-processing step to limit each defect label to its exact atomic position as shown for the dangling bond defect (in blue) near the bottom-right corner of \reffig{defect_lattice_figure}.  Since it is necessary to keep some distance from the various defects anyway, such filtering is not a crucial step of the proposed methodology. However, localizing stray dangling bond defects with atomic accuracy can be beneficial in another way: if they happen to coincide with SiDB positions of placed gates, they can be seamlessly integrated rather than considering them as defects that must be avoided.}

Once each lattice position has been classified, the atomic defects of the \hsi surface can be further divided into two categories: \emph{charged} and \emph{uncharged} (neutral).  The overall charge of a defect is dependent on both the atomic structure and the crystal doping level.  Since the crystals considered in this work are degenerately n-doped, all electron energy levels within the band gap are filled resulting in negatively charged defects.  %

These residual negative charges (as observed in dangling bond and silicon vacancy defects) are able to exert screened Coulombic effects on the charge state of nearby SiDBs as demonstrated by experiments in the literature~\cite{huff2019landscape}. Different gates may have a varying tolerance against these effects, which can be found by running fixed charge defect simulations for defect types of interest using \siqad~\cite{ng2020siqad, ng2022arXiv}. The physical parameters of silicon vacancy defects have been fitted in~\cite{huff2019landscape} and recreated in simulation in~\cite{ng2022arXiv}. With these resources, we have developed the following procedure to determine the minimum distance that each gate tile of a given library must avoid a defect by in order to achieve correct logic operation for that tile in isolation:
\begin{enumerate}
	\item Place the defect at distance $d$ from the gate.
	\item Run physical simulations of the gate toggling through all possible input signal combinations.
	\item For each simulation run, check that the input and output SiDB pairs hold the correct binary logic state.
	\item If any input combination results in incorrect logic states, return~$d$ as the minimum avoidance distance.
	\item Repeat for a sufficient count of defect locations and values of~$d$ to cover meaningful distances in the vicinity of the gate.
\end{enumerate}
We have applied this procedure to the \emph{Bestagon} gate library~\cite{walter2022hexagons} with silicon vacancy defects placed at \SI{7.68}{\angstrom} spacings in both~$x$ and~$y$ directions, \SI{4}{\angstrom} under the surface within a $\approx \SI{30}{\nano\meter} \times \SI{30}{\nano\meter}$ grid centered on each gate. We have found that some Bestagon tiles are not functional at any tested defect distance, necessitating the redesign of these nonfunctional logic tiles.
We have, therefore, redesigned them using an automated SiDB layout designer based on reinforcement learning~\cite{lupoiu2022automated} and selected candidates that have the lowest minimum avoidance distances.
The global minimum avoidance distance was found to be \SI{10}{\nano\meter} by taking the worst performing minimum avoidance distance out of all tiles.
A repository is made publicly available on GitHub\footnote{\url{https://github.com/cda-tum/sidb-defect-aware-physical-design}}
containing
\begin{enumerate*}
	\item \siqad design files of the redesigned tiles,
	\item evaluation results for minimum avoidance distance of these tiles,
	\item the original Bestagon tiles, and
	\item a list of nonfunctional tiles that do not have a proposed direct replacement.
\end{enumerate*}
The latter applies foremost to the half-adder tile which can however be decomposed into, \eg, an XOR and an AND tile.

\subsection{Automatic Physical Design} \label{sec:physical_design}

\begin{figure}[!t]
	\centering
	\subfloat[Overlaying a hexagonal surface tiling. Each tile can hold up to one SiDB gate or wire segment.]{
		\includegraphics[width=.45\linewidth]{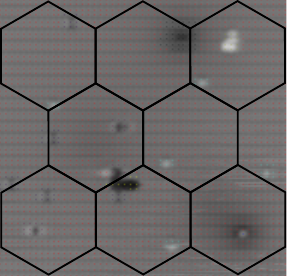}
		\label{fig:defect_tiling:hexagon}
	} \hfil
	\subfloat[Matching Huff~\etal's SiDB OR gate (cf.~\reffig{lattice-and-bdl})~\cite{huff2018binary} against every tile. Crossed-out black dots indicate a conflict with a present atomic defect at that position.]{
		\includegraphics[width=.45\linewidth]{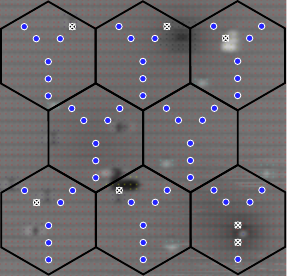}
		\label{fig:defect_tiling:gates}
	}
	\caption{The STM surface scan with atomic defects from \reffig{defects} and \reffig{defect_lattice_figure} with a hexagonal tiling and a gate overlay.}
	\label{fig:defect_tiling}
\end{figure}

The de-facto standard for physical design in the FCN domain is that of a tile-based abstraction~\cite{Huang2005, Blair18, retallick2021low, walter2018exact, walter2021one, walter2019scalable}. That is, a (uniform) tiling of a surface is provided where each tile can implement one designated Boolean function. A standard library of pre-designed SiDB gates and wire segments can be applied to generate a dot-accurate layout from such a gate-level abstraction. Thereby, the focus is shifted from the physical to the logic level and, thus, assists placement and routing by limiting the search space~\cite{walter2019np, walter2018exact, walter2021one, walter2019scalable}. Additionally, \mbox{logic-level} simulation and verification are enabled~\cite{walter2020verify}.

Surface tilings come in manifold forms with the most common one being the Cartesian grid where each tile is a rectangle~\cite{Huang2005, Blair18, Campos16, vankamamidi2006clocking}. Recently, hexagonal tilings were established for SiDBs~\cite{walter2022hexagons}, because they intrinsically match the Y-shaped SiDB gates that have been experimentally proven by Huff \etal~\cite{huff2018binary}, and as such provide a more realistic abstraction for SiDB circuit layouts.

This work, thus, also relies on tile-based design. In the following, we assume a hexagonal tiling together with established SiDB gates~\cite{walter2022hexagons}. However, our approach is generic such that any tiling and any standard library can be applied. 

Instead of imagining a perfect, idealized surface, we consider a realistic STM surface scan as input to our algorithm. We overlay a tiling and match each gate and wire of a given gate library in every rotation against each tile and analyze the effects of defects in the proximity of each dot that make up the gate/wire tile. This procedure yields a blacklist of gate-tile pairs, \ie, a listing of SiDB structures that cannot operate properly on certain tiles of the surface.

We pass this blacklist as a set of placement and routing constraints to a satisfiability-based algorithm that avoids placing those gates/wires on the specified tiles in the specified rotation. The result is a dot-accurate layout that avoids all defects on the surface and, thus, preserves functionality in the presence of disturbances.

\begin{table*}[!t]
	\centering
	\caption{Layout data obtained from physical design on experimentally fabricated and simulated \hsi surfaces.}
	\label{tab:results}
	\begin{minipage}{\linewidth}
		\centering
		\begin{adjustbox}{max width=\linewidth}
			\begin{tabular}{lrrrrrrrrrrrrrrrrrrrrrrrrrr}
				\toprule
				\multicolumn{3}{c}{\multirow{1}{*}{\textsc{Benchmark}~\cite{walter2022hexagons}}} &   & \multicolumn{23}{c}{\textsc{Examined Surface Data}} \\
				\cmidrule(lr){1-3} \cmidrule(lr){5-27}
				\multicolumn{3}{c}{~} & &  \multicolumn{5}{c}{\textsc{Experimental STM Scans}} &  & \multicolumn{8}{c}{\textsc{Simulated w/ Charged Defects}} &  & \multicolumn{8}{c}{\textsc{Simulated w/o Charged Defects}} \\
				\cmidrule(lr){5-9} \cmidrule(lr){11-18} \cmidrule(lr){20-27}
				\multicolumn{1}{c}{~} & \multicolumn{2}{c}{defect-free} & &  \multicolumn{2}{c}{\SI{8.57}{\%} defective} &  & \multicolumn{2}{c}{\SI{6.26}{\%} defective} & & \multicolumn{2}{c}{\SI{1}{\%} defective} & & \multicolumn{2}{c}{\SI{0.5}{\%} defective} & & \multicolumn{2}{c}{\SI{0.1}{\%} defective} & & \multicolumn{2}{c}{\SI{1}{\%} defective} & & \multicolumn{2}{c}{\SI{0.5}{\%} defective} & & \multicolumn{2}{c}{\SI{0.1}{\%} defective}  \\
				\cmidrule(lr){2-3} \cmidrule(lr){5-6} \cmidrule(lr){8-9} \cmidrule(lr){11-12} \cmidrule(lr){14-15} \cmidrule(lr){17-18} \cmidrule(lr){20-21} \cmidrule(lr){23-24} \cmidrule(lr){26-27}
				Name & \#SiDBs & A[\si{\square\nm}] & & \#SiDBs & A[\si{\square\nm}] & & \#SiDBs & A[\si{\square\nm}] & & \#SiDBs & A[\si{\square\nm}] & & \#SiDBs & A[\si{\square\nm}] & & \#SiDBs & A[\si{\square\nm}] & & \#SiDBs & A[\si{\square\nm}] & & \#SiDBs & A[\si{\square\nm}] & & \#SiDBs & A[\si{\square\nm}] \\
				\midrule
				xor2            & \SI{59}{}  & \SI{979.55}{}   & &  --- & --- & & ---  & ---  & & \SI{58}{} & \SI{1120.67}{} & & \SI{59}{}   & \SI{979.55}{}  & & \SI{58}{}  & \SI{1120.67}{}  &  & \SI{59}{}  & \SI{1268.12}{} & & \SI{59}{}  & \SI{979.55}{}   & & \SI{59}{}   & \SI{979.55}{}    \\
				xnor2           & \SI{63}{}  & \SI{979.55}{}   & &  --- & --- & & ---  & ---  & & \SI{62}{} & \SI{1120.67}{} & & \SI{63}{}   & \SI{979.55}{}  & & \SI{62}{}  & \SI{1120.67}{}  &  & \SI{63}{}  & \SI{1268.12}{} & & \SI{63}{}  & \SI{979.55}{}   & & \SI{63}{}   & \SI{979.55}{}    \\
				par\_gen        & \SI{99}{}  & \SI{1956.15}{}  & &  --- & --- & & ---  & ---  & & ---       & ---            & & \SI{97}{}   & \SI{2094.47}{} & & \SI{111}{} & \SI{1882.72}{}  &  & \SI{98}{}  & \SI{1898.50}{} & & \SI{98}{}  & \SI{1956.15}{}  & & \SI{98}{}   & \SI{1956.15}{}   \\
				mux21           & \SI{177}{} & \SI{3447.67}{}  & &  --- & --- & & ---  & ---  & & ---       & ---            & & ---         & ---            & & \SI{230}{} & \SI{5540.22}{}  &  & ---        & ---            & & \SI{163}{} & \SI{3842.41}{}  & & \SI{179}{}  & \SI{5533.14}{}   \\
				par\_check      & \SI{317}{} & \SI{6051.59}{}  & &  --- & --- & & ---  & ---  & & ---       & ---            & & ---         & ---            & & \SI{358}{} & \SI{7924.58}{}  &  & ---        & ---            & & \SI{382}{} & \SI{20899.23}{} & & \SI{229}{}  & \SI{6577.13}{}   \\
				xor5\_r1        & \SI{200}{} & \SI{3447.67}{}  & &  --- & --- & & ---  & ---  & & ---       & ---            & & \SI{216}{}  & \SI{7524.68}{} & & \SI{174}{} & \SI{4682.02}{}  &  & ---        & ---            & & \SI{198}{} & \SI{5629.58}{}  & & \SI{210}{}  & \SI{3941.50}{}   \\
				xor5\_majority  & \SI{191}{} & \SI{3445.16}{}  & &  --- & --- & & ---  & ---  & & ---       & ---            & & \SI{209}{}  & \SI{7524.68}{} & & \SI{166}{} & \SI{4682.02}{}  &  & \SI{181}{} & \SI{5533.14}{} & & \SI{205}{} & \SI{5629.58}{}  & & \SI{257}{}  & \SI{5926.85}{}   \\
				t               & \SI{459}{} & \SI{7924.58}{}  & &  --- & --- & & ---  & ---  & & ---       & ---            & & ---         & ---            & & \SI{443}{} & \SI{8724.82}{}  &  & ---        & ---            & & \SI{541}{} & \SI{20502.28}{} & & \SI{424}{}  & \SI{9768.67}{}   \\
				t\_5            & \SI{482}{} & \SI{7924.58}{}  & &  --- & --- & & ---  & ---  & & ---       & ---            & & ---         & ---            & & \SI{458}{} & \SI{8941.14}{}  &  & ---        & ---            & & \SI{487}{} & \SI{17122.00}{} & & \SI{436}{}  & \SI{10416.59}{}  \\
				c17             & \SI{341}{} & \SI{6330.29}{}  & &  --- & --- & & ---  & ---  & & ---       & ---            & & ---         & ---            & & \SI{391}{} & \SI{7924.58}{}  &  & ---        & ---            & & \SI{598}{} & \SI{20577.78}{} & & \SI{466}{}  & \SI{10316.02}{}  \\
				majority        & \SI{545}{} & \SI{10445.78}{} & &  --- & --- & & ---  & ---  & & ---       & ---            & & ---         & ---            & & \SI{538}{} & \SI{15306.82}{} &  & ---        & ---            & &  ---       & ---             & & \SI{665}{}  & \SI{14301.76}{}  \\
				majority\_5\_r1 & \SI{509}{} & \SI{9518.88}{}  & &  --- & --- & & ---  & ---  & & ---       & ---            & & ---         & ---            & & \SI{570}{} & \SI{11743.40}{} &  & ---        & ---            & &  ---       & ---             & & \SI{925}{}  & \SI{17490.35}{}  \\
				cm82a\_5        & \SI{928}{} & \SI{17502.73}{} & &  --- & --- & & ---  & ---  & & ---       & ---            & & ---         & ---            & &  ---       &  ---            &  & ---        & ---            & &  ---       & ---             & & \SI{1579}{} & \SI{31776.18}{}  \\
				newtag          & \SI{511}{} & \SI{9518.88}{}  & &  --- & --- & & ---  & ---  & & ---       & ---            & & ---         & ---            & & \SI{568}{} & \SI{15306.82}{} &  & ---        & ---            & &  ---       & ---             & & \SI{665}{}  & \SI{15098.90}{}  \\
				\bottomrule
			\end{tabular}
		\end{adjustbox}
	\end{minipage}
\end{table*}

The following example shall illustrate this process.
Assume the STM surface scan depicted in \reffig{defects} was to be used as input to the proposed defect-aware physical design algorithm together with Huff \etal's SiDB OR gate~\cite{huff2018binary} as the target technology. The detected defects are to be avoided while employing a tile-based design paradigm for abstraction and search space restriction. \reffig{defect_tiling:hexagon} shows a hexagonal tiling laid over said STM surface scan from \reffig{defects}. Note that the size and rotation of the hexagons depend on the applied standard library.

In \reffig{defect_tiling:gates}, Huff \etal's OR gate is matched against each tile and the effects of present defects on each SiDB are analyzed. The SiDBs indicated with a crosed-out black dot %
conflict with surface defects. Any gate that has at least one conflicting SiDB is excluded from being placed on that particular tile in that particular rotation. The same is repeated for all rotations of all gates/wires in the given standard library. This procedure yields the aforementioned placement blacklist.

For any given circuit specification, the resulting placement and routing problem with the blacklist is encoded as a satisfiability instance as successfully demonstrated in~\cite{walter2018exact, walter2021one}. A satisfying solution to the instance yields a conflict-free placement and routing on the surface. Due to the search space restrictions that have been employed by relying on tile-based design, it cannot be guaranteed that no such solution exists if the satisfiability instance returns \texttt{UNSAT}. However, the specification can be resynthesized and/or the tile overlay shifted across the surface in order to attempt finding a satisfying solution. 

\section{Experimental Evaluations} \label{sec:eval}

In this section, we present and discuss the results of an experimental evaluation of the proposed defect-aware physical design approach. To this end, we applied the algorithm to automatically generate SiDB layouts on defective \hsi surfaces, both real and simulated. The following \refsec{setup} goes over our experimental setups while \refsec{data} discusses the results and their implications.

\subsection{Experimental Setups} \label{sec:setup}

\subsubsection{Fabrication}

Of the surfaces used in this evaluation, two were experimentally fabricated in a lab and measured with an STM, the others were simulated based on experimental findings.

\sisetup{per-mode=fraction}

The STM measurements were performed using an Omicron LT-STM system operating at \SI{4.5}{\kelvin} and ultra-high vacuum (\SI{3e-11}{\torr}).  The STM tips were electrochemically etched from tungsten wire and sharpened using a field ion microscope~\cite{Rezeq2006fim}.  The used samples are highly arsenic-doped ($\approx \SI{1.5e19}{\atoms\per\cubic\centi\meter}$).  They were prepared \emph{in-situ} via resistive heating.  To this end, they were first degassed at \SI{600}{\degreeCelsius} overnight followed by multiple flash annealing cycles at \SI{1250}{\degreeCelsius}. Finally, the samples were \mbox{hydrogen-terminated} at \SI{330}{\degreeCelsius} while exposing their surface to molecular hydrogen~(\SI{e6}{\torr}).  The H$_{2}$ gas was converted to atomic hydrogen using a tungsten filament held at \SI{1600}{\degreeCelsius}.

The image acquisition was done using a Nanonis SPM controller with respective software.  All images were taken in constant height mode with an imaging bias of \SI{1.3}{\volt} and a current setpoint of \SI{50}{\pico\ampere}.  %

\subsubsection{Programming}

The architecture and training of the neural network used for defect identification in STM scans is modeled after~\cite{rashidi2020deep} and implemented in Python using Keras with the \mbox{TensorFlow} backend. As an addition, the training data was expanded by a factor of three and the number of classes increased to a total of~13 different defect types. The proposed defect-aware physical design algorithm was implemented in C++17 on top of the \fiction framework~\cite{fiction} as part of the \emph{Munich Nanotech Toolkit}~(MNT).\footnote{Publicly available at \url{https://github.com/cda-tum/fiction}.} %
The utilized SMT solver is \emph{Z3}~\cite{moura2008z3}. All experiments were compared against the state-of-the-art results for defect-free SiDB layouts presented in~\cite{walter2022hexagons}. %
The obtained layouts were formally verified for logical correctness using the approach presented in~\cite{walter2020verify}. All evaluations were run on a Manjaro~23 machine with an AMD Ryzen~7~PRO 5850U CPU with 1.90\,GHz (up to 4.40\,GHz boost) and 32\,GB of DDR4 main memory.

\subsection{Results} \label{sec:data}

The STM scans of the fabricated \hsi surfaces span a total of $830 \times 652$ and $740 \times 1090$ hydrogen sites, respectively, of which \SI{8.57}{\%} and \SI{6.26}{\%} are defective.\footnote{Not counting stray DB defects, as they can be erased from the surface prior to the fabrication of circuitry~\cite{huff2017atomic}.} We applied the proposed atomic defect-aware physical design algorithm to generate the same set of benchmark circuits used in~\cite{walter2022hexagons} while obeying the presented surface defect model, and using the \emph{Bestagon} gates that we redesigned for defect robustness.

The results for these cases are listed in \reftab{results} under the caption \textsc{Experimental STM Scans}. For all evaluations, we list the number of required SiDBs to implement each circuit and its bounding box area in \SI{}{\square\nm}. As can be seen, not a single layout could be successfully generated---as is indicated by the dashes---due to the relatively high defect rates of the fabricated samples---highlighting the critical severity of atomic defects in logic design.

To this end, we strive for quantifying their impact by evaluating simulated surfaces of comparable size with variable defect rates of \SI{1}{\%}, \SI{0.5}{\%}, and \SI{0.1}{\%}, once with both charged and neutral atomic defects, and once with only neutral defects, using the same benchmark set. The obtained results can be found in the same table under the caption \textsc{Simulated w/ Charged Defects} for surfaces including charged defects, and under \textsc{Simulated w/o Charged Defects} for surfaces with purely neutral defects. In both cases, defect types were automatically distributed in accordance with experimental findings. In the former case, charged ones make up \SI{5}{\%} of all defects.

Three core findings can be obtained from these results: \begin{enumerate*}
	\item defect avoidance directly correlates with significantly larger overall area consumption,
	\item high defect rates, and particularly charged defects, have a tremendous impact on layout generation to the degree where circuits cannot be realized in their vicinity at all, and
	\item fabrication capabilities must achieve a defect rate of around \SI{0.1}{\%} in the absence of charged defects or $< \SI{0.1}{\%}$ with charged defects present to enable sophisticated layout manufacturing.
\end{enumerate*}

\section{Conclusions} \label{sec:concl}

Fabrication capabilities of \emph{Silicon Dangling Bonds}~(SiDBs) have advanced to the automated manufacturing stage. Nevertheless, atomic substrate defects are currently preventing technology scaling as they disturb gate functionalities or prevent logic realization altogether. In this work, we presented a surface defect model to guide physical design that we obtained by investigating 13 experimentally verified \hsi defect types. Furthermore, we proposed modifications to an established SiDB gate library to increase its robustness against substrate defects. Finally, we proposed a defect-aware placement and routing algorithm that considers STM surface scans obtained from experimentation as well as simulated surface data and designs functioning SiDB circuit layouts in the presence of atomic defects. An experimental evaluation on real fabricated surfaces demonstrated its functioning but also highlighted the limitations of current fabrication capabilities. We demonstrated the critical impact that charged defects have on the creation of circuit layouts and determined a defect rate of around \SI{0.1}{\%}, if charged defects can be completely eliminated, or $< \SI{0.1}{\%}$, otherwise, to be required for the future of large-scale SiDB logic manufacturing. Herewith, this work represents an amalgamation of fabrication and design automation that provides the basis for large-scale defect-aware physical design of SiDB circuitry.

\bibliographystyle{IEEEtran}
\bibliography{./bib/IEEEabrv, ./bib/Bibliography.bib}

\begin{thebibliography}{10}
\providecommand{\url}[1]{#1}
\csname url@samestyle\endcsname
\providecommand{\newblock}{\relax}
\providecommand{\bibinfo}[2]{#2}
\providecommand{\BIBentrySTDinterwordspacing}{\spaceskip=0pt\relax}
\providecommand{\BIBentryALTinterwordstretchfactor}{4}
\providecommand{\BIBentryALTinterwordspacing}{\spaceskip=\fontdimen2\font plus
\BIBentryALTinterwordstretchfactor\fontdimen3\font minus
  \fontdimen4\font\relax}
\providecommand{\BIBforeignlanguage}[2]{{%
\expandafter\ifx\csname l@#1\endcsname\relax
\typeout{** WARNING: IEEEtran.bst: No hyphenation pattern has been}%
\typeout{** loaded for the language `#1'. Using the pattern for}%
\typeout{** the default language instead.}%
\else
\language=\csname l@#1\endcsname
\fi
#2}}
\providecommand{\BIBdecl}{\relax}
\BIBdecl

\bibitem{haider2009controlled}
M.~B. Haider \emph{et~al.}, ``{Controlled Coupling and Occupation of Silicon
  Atomic Quantum Dots at Room Temperature},'' \emph{Physical Review Letters},
  vol. 102, no.~4, p. 046805, 2009.

\bibitem{huff2017atomic}
T.~R. Huff \emph{et~al.}, ``{Atomic White-Out: Enabling Atomic Circuitry
  through Mechanically Induced Bonding of Single Hydrogen Atoms to a Silicon
  Surface},'' \emph{ACS Nano}, vol.~11, no.~9, pp. 8636--8642, Sep. 2017.

\bibitem{pavlicek2017tip}
N.~Pavli\u{c}ek \emph{et~al.}, ``{Tip-induced Passivation of Dangling Bonds on
  Hydrogenated {S}i(100)-2$\times$1},'' \emph{Applied Physics Letters}, vol.
  111, no.~5, p. 053104, 2017.

\bibitem{achal2018lithography}
R.~Achal \emph{et~al.}, ``{Lithography for Robust and Editable Atomic-Scale
  Silicon Devices and Memories},'' \emph{Nature Communications}, vol.~9, no.~1,
  p. 2778, Jul. 2018.

\bibitem{Wolkow14}
R.~A. Wolkow \emph{et~al.}, \emph{{Silicon Atomic Quantum Dots Enable
  Beyond-CMOS Electronics}}.\hskip 1em plus 0.5em minus 0.4em\relax Springer,
  2014, pp. 33--58.

\bibitem{D0NR08295C}
\BIBentryALTinterwordspacing
J.~Croshaw \emph{et~al.}, ``{Ionic Charge Distributions in Silicon Atomic
  Surface Wires},'' \emph{Nanoscale}, vol.~13, pp. 3237--3245, 2021. [Online].
  Available: \url{http://dx.doi.org/10.1039/D0NR08295C}
\BIBentrySTDinterwordspacing

\bibitem{wyrick2019atom}
J.~Wyrick \emph{et~al.}, ``{Atom-by-Atom Fabrication of Single and Few Dopant
  Quantum Devices},'' \emph{Advanced Functional Materials}, vol.~29, p.
  1903475, 2019.

\bibitem{huff2018binary}
T.~Huff \emph{et~al.}, ``{Binary Atomic Silicon Logic},'' \emph{Nature
  Electronics}, vol.~1, pp. 636--643, 2018.

\bibitem{Anderson14}
N.~G. Anderson \emph{et~al.}, \emph{Field-coupled Nanocomputing: Paradigms,
  Progress, and Perspectives}.\hskip 1em plus 0.5em minus 0.4em\relax New York:
  Springer, 2014.

\bibitem{jiao2003building}
J.~Jiao \emph{et~al.}, ``{Building Blocks for the Molecular Expression of
  Quantum Cellular Automata. Isolation and Characterization of a Covalently
  Bonded Square Array of Two Ferrocenium and Two Ferrocene Complexes},''
  \emph{Journal of the American Chemical Society}, vol. 125, no.~25, pp.
  7522--7523, 2003.

\bibitem{qi2003molecular}
H.~Qi \emph{et~al.}, ``{Molecular Quantum Cellular Automata Cells. Electric
  Field Driven Switching of a Silicon Surface Bound Array of Vertically
  Oriented Two-Dot Molecular Quantum Cellular Automata},'' \emph{Journal of the
  American Chemical Society}, vol. 125, no.~49, pp. 15\,250--15\,259, 2003.

\bibitem{landauer1961irreversibility}
R.~Landauer, ``{Irreversibility and Heat Generation in the Computing
  Process},'' \emph{IBM Journal of Research and Development}, vol.~5, no.~3,
  pp. 183--191, 1961.

\bibitem{keyes1970minimal}
R.~W. Keyes \emph{et~al.}, ``{Minimal Energy Dissipation in Logic},'' \emph{IBM
  Journal of Research and Development}, vol.~14, no.~2, pp. 152--157, 1970.

\bibitem{lent2006bennett}
C.~S. Lent \emph{et~al.}, ``{Bennett clocking of quantum-dot cellular automata
  and the limits to binary logic scaling},'' \emph{Nanotechnology}, vol.~17,
  no.~16, pp. 4240--4251, 2006.

\bibitem{toth1999quasiadiabatic}
G.~Toth \emph{et~al.}, ``{Quasiadiabatic switching for metal-island quantum-dot
  cellular automata},'' \emph{Journal of Applied Physics}, vol.~85, no.~5, pp.
  2977--2984, 1999.

\bibitem{Timler02}
J.~Timler \emph{et~al.}, ``{Power Gain and Dissipation in {Q}uantum-dot
  {C}ellular {A}utomata},'' \emph{Journal of Applied Physics}, vol.~91, no.~2,
  pp. 823--831, 2002.

\bibitem{livadaru2010danglingbond}
L.~Livadaru \emph{et~al.}, ``{Dangling-bond Charge Qubit on a Silicon
  Surface},'' \emph{New Journal of Physics}, vol.~12, no.~8, p. 083018, Aug.
  2010.

\bibitem{ng2020thes}
S.~S.~H. Ng, ``{Computer-aided Design of Atomic Silicon Quantum Dots and
  Computational Applications},'' Master's thesis, University of British
  Columbia, 2020.

\bibitem{chiu2020poissolver}
H.~N. Chiu \emph{et~al.}, ``{{{PoisSolver}}: A Tool for Modelling Silicon
  Dangling Bond Clocking Networks},'' in \emph{IEEE-NANO}.\hskip 1em plus 0.5em
  minus 0.4em\relax {Montreal, QC, Canada}: {IEEE}, Jul. 2020, pp. 134--139.

\bibitem{chiu2020thes}
H.~N. Chiu, ``{Simulation and Analysis of Clocking and Control for
  Field-coupled Quantum-dot Nanostructures},'' Master's thesis, University of
  British Columbia, 2020.

\bibitem{ng2020siqad}
S.~S.~H. Ng \emph{et~al.}, ``{SiQAD: A Design and Simulation Tool for Atomic
  Silicon Quantum Dot Circuits},'' \emph{TNANO}, vol.~19, pp. 137--146, 2020.

\bibitem{drewniok2023quicksimIEEE}
J.~Drewniok \emph{et~al.}, ``{\emph{QuickSim}: Efficient \emph{and} Accurate
  Physical Simulation of Silicon Dangling Bond Logic},'' in \emph{IEEE-NANO},
  2023, pp. 817--822.

\bibitem{drewniok2023temperatureIEEE}
------, ``{Temperature Behavior of Silicon Dangling Bond Logic},'' in
  \emph{IEEE-NANO}, 2023, pp. 925--930.

\bibitem{drewniok2023need}
------, ``The need for speed: Efficient exact simulation of silicon dangling
  bond logic,'' 2023.

\bibitem{bahar2020atomic}
A.~N. Bahar \emph{et~al.}, ``{Atomic Silicon Quantum Dot: A New Designing
  Paradigm of an Atomic Logic Circuit},'' \emph{TNANO}, pp. 807--810, 2020.

\bibitem{vieira2021novel}
M.~D. Vieira \emph{et~al.}, ``{Novel Three-Input Gates for Silicon Quantum
  Dot},'' in \emph{SBCCI}, 2021, pp. 1--6.

\bibitem{vieira2022three}
------, ``{Three-Input NPN Class Gate Library for Atomic Silicon Quantum
  Dots},'' \emph{IEEE Design \& Test}, 2022.

\bibitem{walter2022hexagons}
M.~Walter \emph{et~al.}, ``{Hexagons are the Bestagons: Design Automation for
  Silicon Dangling Bond Logic},'' in \emph{DAC}, vol.~22, 2022.

\bibitem{ng2023ablueprint}
S.~S.~H. Ng \emph{et~al.}, ``{A Blueprint for Machine Learning Accelerators
  Using Silicon Dangling Bonds},'' in \emph{IEEE-NANO}, 2023.

\bibitem{hofmann2023hexagonalization}
S.~Hofmann \emph{et~al.}, ``{Scalable Physical Design for Silicon Dangling Bond
  Logic: How a 45\textdegree~Turn Prevents the Reinvention of the Wheel},'' in
  \emph{IEEE-NANO}, 2023, pp. 872--877.

\bibitem{boland1993}
J.~J. Boland, ``{Scanning tunnelling microscopy of the interaction of hydrogen
  with silicon surfaces},'' \emph{Advances in Physics}, vol.~42, no.~2, pp.
  129--171, 1993.

\bibitem{rashidi2020deep}
M.~Rashidi \emph{et~al.}, ``{Deep learning-guided surface characterization for
  autonomous hydrogen lithography},'' \emph{Machine Learning: Science and
  Technology}, vol.~1, no.~2, 2020.

\bibitem{croshaw2020atomic}
J.~Croshaw \emph{et~al.}, ``{Atomic defect classification of the {{H-Si(100)}}
  surface through multi-mode scanning probe microscopy},'' \emph{Beilstein
  Journal of Nanotechnology}, vol.~11, no.~1, pp. 1346--1360, 2020.

\bibitem{huff2019landscape}
T.~Huff \emph{et~al.}, ``{Electrostatic Landscape of a Hydrogen-Terminated
  Silicon Surface Probed by a Moveable Quantum Dot},'' \emph{ACS Nano},
  vol.~13, no.~9, pp. 10\,566--10\,575, 2019.

\bibitem{rashidi2022automated}
M.~Rashidi \emph{et~al.}, ``{Automated Atomic Scale Fabrication},'' US Patent
  20\,220\,130\,033, 2022.

\bibitem{huff2019electrostatic}
T.~Huff \emph{et~al.}, ``{Electrostatic Landscape of a Hydrogen-Terminated
  Silicon Surface Probed by a Moveable Quantum Dot},'' \emph{ACS Nano},
  vol.~13, no.~9, pp. 10\,566--10\,575, 2019.

\bibitem{achal2019detecting}
R.~Achal \emph{et~al.}, ``{Detecting and Directing Single Molecule Binding
  Events on H-Si (100) with Application to Ultradense Data Storage},''
  \emph{ACS Nano}, vol.~14, no.~3, pp. 2947--2955, 2019.

\bibitem{onoda2021ohmic}
J.~Onoda \emph{et~al.}, ``{Ohmic Contact to Two-Dimensional Nanofabricated
  Silicon Structures with a Two-Probe Scanning Tunneling Microscope},''
  \emph{ACS Nano}, vol.~15, no.~12, pp. 19\,377--19\,386, 2021.

\bibitem{altincicek2022atomically}
F.~Altincicek, ``{Atomically Defined Wires on P-Type Silicon},'' \emph{Bulletin
  of the American Physical Society}, 2022.

\bibitem{pitters2011charge}
J.~L. Pitters \emph{et~al.}, ``{Charge Control of Surface Dangling Bonds Using
  Nanoscale Schottky Contacts},'' \emph{ACS Nano}, vol.~5, no.~3, pp.
  1984--1989, Mar. 2011.

\bibitem{rashidi2018initiating}
M.~Rashidi \emph{et~al.}, ``Initiating and monitoring the evolution of single
  electrons within atom-defined structures,'' \emph{Physical Review Letters},
  vol. 121, no.~16, p. 166801, Oct. 2018.

\bibitem{rashidi2016timeresolved}
------, ``{Time-resolved single dopant charge dynamics in silicon},''
  \emph{Nature Communications}, vol.~7, no.~1, p. 13258, Dec. 2016.

\bibitem{Wolkow2021}
R.~A. Wolkow \emph{et~al.}, ``{Multiple silicon atom quantum dot and devices
  inclusive thereof},'' US Patent 10\,937\,959, 2021.

\bibitem{Wolkow2021a}
R.~Wolkow \emph{et~al.}, ``{Initiating and monitoring the evolution of single
  electrons within atom-defined structures},'' US Patent 11\,047\,877, 2021.

\bibitem{Boland1992}
J.~J. Boland, ``{Role of bond-strain in the chemistry of hydrogen on the
  Si(100) surface},'' \emph{Surface Science}, vol. 261, no. 1-3, pp. 17--28,
  1992.

\bibitem{ng2022arXiv}
\BIBentryALTinterwordspacing
S.~S.~H. Ng \emph{et~al.}, ``{Charged Defect Simulation in SiDB Systems},''
  2022. [Online]. Available: \url{https://arxiv.org/abs/2211.08698}
\BIBentrySTDinterwordspacing

\bibitem{lupoiu2022automated}
R.~Lupoiu \emph{et~al.}, ``{Automated Atomic Silicon Quantum Dot Circuit Design
  via Deep Reinforcement Learning},'' 2022.

\bibitem{Huang2005}
J.~Huang \emph{et~al.}, ``{Tile-based {QCA} Design Using Majority-like Logic
  Primitives},'' \emph{JETC}, vol.~1, no.~3, pp. 163--185, 2005.

\bibitem{Blair18}
E.~Blair and C.~Lent, ``{Clock Topologies for Molecular Quantum-Dot Cellular
  Automata},'' \emph{Journal of Low Power Electronics and Applications},
  vol.~8, no.~3, 2018.

\bibitem{retallick2021low}
J.~Retallick \emph{et~al.}, ``{Low-Energy Eigenspectrum Decomposition (LEED) of
  Quantum-Dot Cellular Automata Networks},'' \emph{TNANO}, vol.~20, pp.
  104--112, 2021.

\bibitem{walter2018exact}
M.~Walter \emph{et~al.}, ``{An Exact Method for Design Exploration of
  {Q}uantum-dot {C}ellular {A}utomata},'' in \emph{DATE}, 2018, pp. 503--508.

\bibitem{walter2021one}
------, ``{One-pass Synthesis for Field-coupled Nanocomputing Technologies},''
  in \emph{ASP-DAC}.\hskip 1em plus 0.5em minus 0.4em\relax ACM New York, NY,
  USA, 2021, pp. 574--580.

\bibitem{walter2019scalable}
------, ``{Scalable Design for Field-coupled Nanocomputing Circuits},'' in
  \emph{ASP-DAC}.\hskip 1em plus 0.5em minus 0.4em\relax ACM New York, NY, USA,
  2019, pp. 197--202.

\bibitem{walter2019np}
------, ``{Placement \& Routing for Tile-based Field-coupled Nanocomputing
  Circuits is $\mathcal{N\!P}$-complete},'' \emph{JETC}, vol.~15, no.~3, 2019.

\bibitem{walter2020verify}
------, ``{Verification for Field-coupled Nanocomputing Circuits},'' in
  \emph{DAC}, 2020.

\bibitem{Campos16}
C.~A.~T. Campos \emph{et~al.}, ``{USE: A Universal, Scalable, and Efficient
  Clocking Scheme for QCA},'' \emph{TCAD}, vol.~35, no.~3, pp. 513--517, 2016.

\bibitem{vankamamidi2006clocking}
V.~Vankamamidi \emph{et~al.}, ``{Clocking and Cell Placement for QCA},'' in
  \emph{IEEE-NANO}, vol.~1.\hskip 1em plus 0.5em minus 0.4em\relax IEEE, 2006,
  pp. 343--346.

\bibitem{Rezeq2006fim}
M.~Rezeq, J.~Pitters, and R.~Wolkow, ``{Tungsten nanotip fabrication by
  spatially controlled field-assisted reaction with nitrogen},'' \emph{Journal
  of Chemical Physics}, vol. 124, no. 204716, 2006.

\bibitem{fiction}
M.~Walter \emph{et~al.}, ``{fiction: An Open Source Framework for the Design of
  Field-coupled Nanocomputing Circuits},'' 2019.

\bibitem{moura2008z3}
L.~de~Moura, ``{Z3: An efficient SMT solver},'' in \emph{International
  Conference on Tools and Algorithms for the Construction and Analysis of
  Systems}.\hskip 1em plus 0.5em minus 0.4em\relax Springer, 2008, pp.
  337--340.

\end{thebibliography}

\balance

\end{document}